\documentclass{mem}
\usepackage{natbib}
\usepackage{txfonts}
\usepackage{balance}
\usepackage{graphicx}
\usepackage{flushend}
\usepackage{xcolor}
\usepackage[breaklinks,pdftex]{hyperref}
\idline{2}{1}
\begin{document}
\def\teff{$T\rm_{eff }$}
\def\kms{$\mathrm {km s}^{-1}$}
\newcommand{\stef}{\textcolor{blue}}
\newcommand{\asa}{\textcolor{teal}}
\newcommand{\asacom}[1]{\color{teal}[ÁS: #1]\color{black}}
\newcommand{\ire}{}

\title{
Are all metal-poor stars of second-generation?
}

   \subtitle{}

\author{
I. \,Vanni\inst{1, 2}, 
S. \,Salvadori\inst{1,2}
\and Á. \,Skúladóttir\inst{1,2} 
          }

\institute{
Dipartimento di Fisica e Astrofisica, Università degli Studi di Firenze, Via G. Sansone 1, I-50019, Sesto Fiorentino, Italy
\and
INAF/Osservatorio Astrofisico di Arcetri, Largo E.Fermi 5, I-50125, Firenze, Italy \\
\email{irene.vanni@unifi.it}
}

\authorrunning{Vanni I.}

\titlerunning{Are all metal-poor stars of second-generation?}

\date{Received: Day Month Year; Accepted: Day Month Year}

\abstract{
Hydrodynamical cosmological simulations predict that the metal-free Population~III (Pop~III) stars were likely very massive and, therefore, short-lived. However, they left their chemical imprint on their descendants, which can also have masses $ < 0.8 \mathrm {M_{\odot}}$ and still be alive today. The Milky Way stellar halo is one of the oldest and most metal-poor component of the Local Group and a peculiar class of stars, the so-called Carbon-Enhanced Metal-Poor (CEMP-no) stars %[Fe/H] $<$ -1, [C/Fe] $>$ 0.7), 
seem to be directly related to Pop~III stars. We aim at revealing if \emph{all} metal-poor halo stars are true \emph{second-generation stars} or if they have also been enriched by the subsequent generations of normal (Pop~II) stars. For this purpose, we compare the measured carbon and iron abundances of the metal-poor halo stars with the ones predicted by our simple parametric model, varying the pollution level from Pop~III and normal stars. We find that only the most C-enhanced and Fe-poor stars enclose in their photospheres the pure imprint of Pop~III stars, while, as the [C/Fe] decreases, the probability of being also polluted by normal Pop~II stars increases.
}
\maketitle{}

\section{Introduction}
The Milky Way stellar halo is a promising region to search for the descendants of the first (Pop~III) stars. Here the oldest and most metal-poor stars have been spotted, suggesting a link with the chemical elements produced by Pop~III stars. The mass of Pop~III stars formed in state-of-the-art hydro-dynamical simulations cover a wide mass range, $0.1 <m_{PopIII} < 1000 \mathrm M_{\odot}$ \citep[e.g][]{Hirano2014,Greif2015,Hirano2017}. However, there is general theoretical consensus supporting the idea that the first stars were typically more massive than those forming today, with a characteristic mass $\geq$ 10$\mathrm M_{\odot}$ \citep[e.g.][]{Bromm2013}. Furthermore, a zero-metallicity star has never been observed, confirming the massive nature of Pop~III stars \citep[e.g.][]{Rossi2021}. If Pop~III stars were typically massive, most of them would have ended their lives in less than 100 Myrs, exploding as supernovae (SNe). Thus, the first SNe polluted the Inter-Stellar Medium (ISM) with their newly synthesized chemical elements. Consequently, normal Population~II (Pop~II) stars were able to form in this Pop~III-enriched ISM \citep[e.g.][]{Schneider2012}. Among these ``second-generation" stars, those with masses $< 0.8 \mathrm M_{\odot}$ can survive until today preserving in their photospheres the chemical signatures of Pop~III stars.

The iron-abundance\footnote{[Fe/H]=$\log{\left(\frac{N_{Fe}}{N_H}\right)}-\log{\left(\frac{N_{Fe}}{N_H}\right)_{\odot}}$. Solar abundances are from \citet{Asplund2009a}.} of halo stars spans more than 5 dex. Stars with [Fe/H] $< -1$ are called metal-poor and they are separated in two categories depending on their [C/Fe] values \citep[see][]{Beers2005}: Carbon-Enhanced Metal-Poor stars (CEMP) when [C/Fe] $\ge +0.7$, and C-normal otherwise. The CEMP stars can be further divided into CEMP-s(/r) and CEMP-no stars depending on whether [Ba/Fe] is super- or sub-solar, respectively. Neutron-capture elements from the slow process, such as barium, are mainly produced by Asymptotic Giant Branch (AGB) stars and their enhancement in the photosphere of a star is a signature of mass accretion from an AGB companion. The C-excess in CEMP-s stars is thus expected to be acquired during its lifetime, via mass transfer from a binary AGB companion \citep[e.g.][]{Abate2015}. On the other hand, the C-excess in CEMP-no stars is expected to be representative of the environment of formation. This is supported by observations, showing that CEMP-s stars are almost exclusively in binary systems, which is not the case for CEMP-no stars (e.g.~\citealp{HansenT16,Arentsen2019}). 
%{\bf STEF: attention, this is not what David is finding - see todays' talk. }

The fraction of CEMP-no stars increases towards decreasing [Fe/H], suggesting a direct link between the birth clouds of CEMP-no stars and the chemical products of Pop~III SNe \citep[e.g.][]{Salvadori2015,Chiaki2020}. Conversely, the origin of C-normal very metal-poor stars ($\rm[Fe/H]<-2$) is still highly debated: their chemical abundances are indeed consistent with either an enrichment from Pop~III SNe {\it only} \citep[see][]{Hartwig2019,Welsh2020}, or from both Pop~III and normal Pop~II SNe \citep[e.g.][]{deBennassuti2017,Liu2021}. 

In this work, we aim at understanding if all very metal-poor stars are true descendants of Pop~III stars. To this end we will take advantage of the high-resolution measurements of C-enhanced and C-normal metal-poor stars, and compare those with our theoretical models. We refer the reader to Vanni et al. (in prep) for a comparison with {\it all} the chemical abundance ratios measured in metal-poor star.

\section{Sample of metal-poor halo stars}

\label{halo}
The  \emph{halo sample}, which will be used to compare to our model, includes 132 CEMP-no and C-normal halo stars with [Fe/H] $\le -2$, that have been observed with high resolution spectroscopy (R $\ge 30\,000$). In particular, this sample includes the stars presented in \citet{Cayrel2004a} and \citet{Yong2013a}, excluding CEMP-s stars, as well as extremely metal-poor stars, [Fe/H] $\le -3$, from: \citet{Christlieb2004,Norris2007,Caffau2011a,Hansen2014,Keller,Frebel2015,Bonifacio2015a,Li2015,Starkenburg2018,Bonifacio2018,Francois2018,Aguado2019,Ezzeddine2019}. These chemical abundances are not corrected for the non-LTE effects. 

The stars of our {\it halo sample} cover a wide iron range, $\rm -7.1<[Fe/H]<-2$, and their metallicity distribution function peaks at [Fe/H]$\sim -3$ (\ire{see e.g. \citealp{Placco2014c}}).
\begin{figure}
\resizebox{\hsize}{!}{\includegraphics[clip=true]{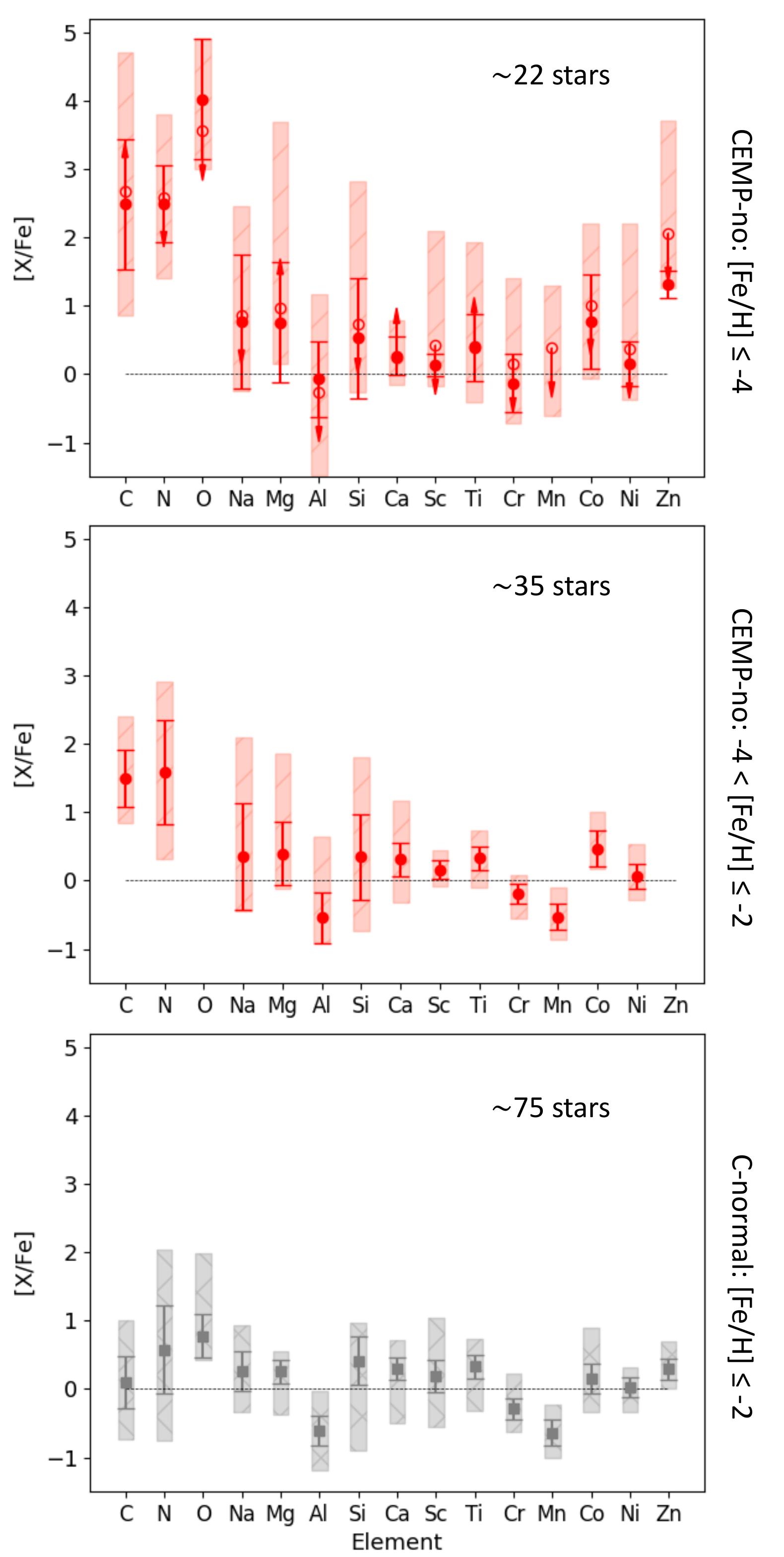}}
    \caption{\footnotesize Chemical abundance pattern of the \emph{halo sample} in three categories: CEMP-no stars with [Fe/H] $\le -4$ (top), CEMP-no with $-4 <$ [Fe/H] $\le -2$ (middle), and C-normal stars with [Fe/H] $\le -2$ (bottom). Filled points with error bars represent the mean and the standard deviation computed without considering upper/lower limits, while the empty points (top panel) with arrows including upper and/or lower limits. Shaded areas show the range of the measured chemical abundances, including also upper/lower limits.}
    \label{fig:Yong_scarti}
\end{figure}
In Fig.~\ref{fig:Yong_scarti} we show the average abundance ratios, standard deviation, and maximum range of the scatter for stars in the {\it halo sample}, divided among: CEMP-no stars (red) with [Fe/H] $< -4$ (upper panel) and with $-4 <$ [Fe/H] $ < -2$ (middle), and finally C-normal stars with $\rm[Fe/H]<-2$ (gray, lower panel). 
%All the stars in the sample have carbon (and iron) abundances while for the other elements the measurements are sometimes limited to upper/lower limits, especially for the most iron-poor stars.
\ire{All the stars in the sample have carbon (and iron) abundance measurements, while for the other elements the measurements are sometimes limited to few stars (e.g. O or Zn, which are not available for the CEMP-no stars in our sample with $-4 <$ [Fe/H] $< -2$). Furthermore, the majority of the most iron-poor stars only have upper/lower limits of abundance values.
For this reason, in the top panel of Fig. \ref{fig:Yong_scarti} we include the mean chemical abundance ratios obtained by both including and excluding the upper/lower limits. 
Furthermore, the upper/lower limits were used to evaluate the range of the scatter. For example, for the C-normal stars we have also used the \cite{Caffau2011a} and \cite{Starkenburg2018} stars which have, respectively, [C/Fe]$<+0.7$ and [C/Fe]$<+1.0$. }

In Fig. \ref{fig:Yong_scarti} we first notice that the standard deviation and scatter in the abundance ratios are highest for the most Fe-poor CEMP-no stars (top panel), while they gradually decrease towards lower panels, being smallest for C-normal stars (see also \citealt{Cayrel2004a}). The mean abundance ratios of the light elements C, N and O, are generally higher for CEMP-no stars than for C-normal stars, even if we consider stars in the same [Fe/H] range. On the other hand, the abundance ratios of the elements heavier than oxygen are consistent within the error bars between the different stellar classes. 

\section{The model}

\label{model}
Here we briefly summarize the simple and general parametric study presented by \citet{Salvadori2019} and further implemented by Vanni et al. (in prep). The model aims to chemically characterize the descendants of Pop~III stars: long-lived stars formed in environments predominantly polluted by Pop~III SNe, i.e. where the metals from Pop~III SNe account for $\geq 50\%$ of metals in the ISM. 

Following the results of hydrodynamical cosmological simulations \cite[e.g.][]{Hirano2014}, we assume that a single Pop~III star can form in the primordial star-forming haloes. Then, we evaluate the chemical enrichment of the ISM after: (i) the injection of heavy elements by Pop~III SNe with different explosion energies and progenitor masses; and (ii) the subsequent contribution of "normal" Pop~II stars exploding as core-collapse SNe. We adopt the yields by \cite{Heger2002a} for very massive Pop~III stars, $m=[140-260]\mathrm M_{\odot}$, that explode as Pair Instability SNe ({\it PISN}, $E_{SN}=10^{52} - 10^{53}$ erg) and by \cite{Heger2010} for intermediate mass Pop~III stars, $m=[10-100]\mathrm M_{\odot}$, exploding as {\it faint}, {\it core-collapse}, {\it high-energy} SNe and {\it hypernovae}, whose explosion energy is respectively equal to $E_{SN}=(0.6,1.2,3.0,10.0)\times 10^{51}$~erg. For Pop~II stars we adopt the yields of \citet{Woosley1995} and \citet{Limongi2018} and show both results to account for the uncertainty due to the choice of the stellar evolutionary model (\ire{see e.g. \citealp{Nomoto2013}}).

The main unknowns related to early cosmic star formation and metal enrichment are encapsulated into the free parameters, which are varied to obtain the most general model. The model takes into consideration: the fraction of gas converted into stars, or the star formation efficiency, $f_*=[10^{-4}-10^{-1}]$; metals retained into the ISM, parameterized with the dilution factor $f_{dil}$; and the mass fraction of metals injected into the ISM by Pop~III stars with respect to the total, $f_{PopIII}=[100-50]\%$.
It turns out that the iron abundance of the ISM depends on $f_{PopIII}$; the yields of Pop~III, $Y^{III}(m,E_{SN})$, and Pop~II stars, $Y^{II}(m,Z)$; and on the ratio between the first two free parameters, $f_*/f_{dil}$. Conversely, [C/Fe] is not affected by $f_*$ nor by $f_{dil}$ (see \citealt{Salvadori2019} for details).

\section{Results}
\label{results}

\begin{figure*}[h]
\resizebox{\hsize}{!}{\includegraphics[clip=true]{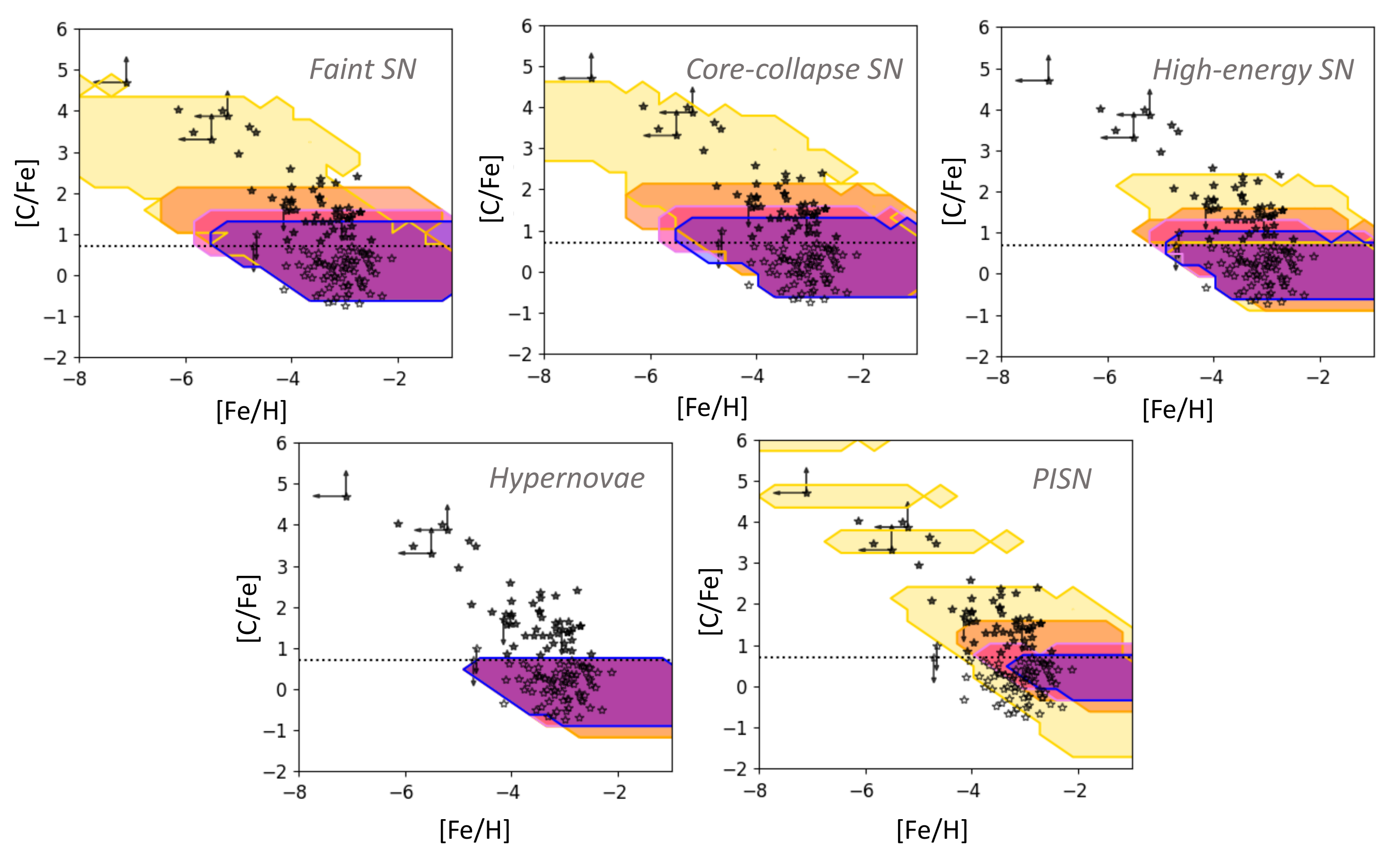}}
\caption{Carbon-to-iron ratios of PopIII descendants. Colours show different level of Pop~III enrichment: 100\% (yellow), 90\% (orange), 70\% (pink), 50\% (purple). Each panel shows different explosion energies of Pop~III SNe (see labels): from faint SNe ($E \sim 6 \times 10^{50}$ erg) to hypernovae ($E \sim 10^{52}$ erg) in the mass range $[10-100] \mathrm M_{\odot}$; as well as PISN ($E = 10^{52-53}$ erg) for $[140-260] \mathrm M_{\odot}$. Star symbols show the measured abundances of CEMP-no (filled) and C-normal stars (open). 
The dotted line is the threshold for the carbon enhancement, $\rm[C/Fe]>+0.7$. \ire{Note that the yellow area for the hypernovae progenitor case is hidden behind the other colours.}}
\label{fig:CsuFe}
\end{figure*}

In Fig. \ref{fig:CsuFe} we compare the predicted [C/Fe] and [Fe/H] values for the descendants of the first stars (shaded areas) with the chemical abundances measured in the \emph{halo sample} (points). The colours denote different fraction of metals from Pop~III SNe: from the yellow area, where the descendants are completely imprinted by Pop~III ($f_{popIII}=100\%$), to the purple area, where they are equally imprinted by Pop~III and Pop~II ($f_{popIII}=50\%$) stars. The different panels show the results for different explosion energies of Pop~III SNe: from the least energetic faint SNe to the most energetic PISNe.

The maximum [C/Fe] that is predicted by the models (Fig.~\ref{fig:CsuFe}) decreases with: (i) increasing explosion energy of Pop~III SNe; and (ii) increasing contribution of Pop~II SNe to the chemical enrichment. Furthermore, we see that CEMP-no halo stars with [C/Fe] $\ge +2.5$ and [Fe/H] $\le -4$ can {\it only} be reproduced by models accounting for a 100\% enrichment from Pop~III stars, which implies that these objects are {\it true second-generation stars}. Indeed, even a 10\% pollution by Pop~II stars is sufficient to lower the maximum to [C/Fe]$<+2.5$.

Which kind of Pop~III SNe imprinted these highly C-enhanced  ultra iron-poor stars? Different Pop~III SN models are able to enrich the ISM with high [C/Fe] and low [Fe/H]: the lightest PISN progenitors, M $\sim 140-150 \mathrm M_{\odot}$, and massive Pop~III stars, $40 \mathrm M_{\odot} \le \mathrm M \le 100 \mathrm M_{\odot}$, exploding as {\it faint} SNe or {\it core-collapse} SNe. Still, our models show that the abundance ratios of the other chemical elements measured in these CEMP-no ultra iron-poor stars (Fig. \ref{fig:Yong_scarti}) are not consistent with a light PISN enrichment, but rather support a scenario where these stars have been imprinted by massive Pop~III stars exploding as low-energy SNe.

We can now try to understand if all metal-poor stars are truly second-generation objects. At higher $\rm[Fe/H]> -4$, the observed properties of \ire{both CEMP-no and C-normal} halo stars can be reproduced by several models (Fig.~\ref{fig:CsuFe}). At $\rm[Fe/H]>-4$, only the hypernovae case is not able to reproduce the [C/Fe] values measured in CEMP-no stars with or without Pop~II contribution. To disentangle this degeneracy of different Pop~III SN models, we would need to study the abundance ratios of other chemical elements. This will be done in a complementary work, Vanni et al. (in prep).

The probability for a star to be uniquely ($100\%$) or predominantly ($>50\%$) imprinted by the chemical products of Pop~III SNe decreases as [C/Fe] decreases. Furthermore, an increasing contribution from Pop~II stars reduces the scatter in [C/Fe] predicted by different models, in agreement to what is observed (Fig.~\ref{fig:Yong_scarti}). Indeed, we can see in Fig. \ref{fig:CsuFe} that the scatter in [C/Fe] spans over $\sim 5$ dex for pure Pop~III star descendants, and only $\sim 2$ dex for the descendants enriched by Pop~III stars at a 50\% level. Thus, the specific chemical features of the ISM enriched by different Pop~III SN types are mostly washed out if Pop~III and Pop~II stars equally contribute to the ISM enrichment. This is a strong indication that a group of stars that show very similar chemical abundances, i.e. a small star-to-star scatter like C-normal halo stars, have likely been polluted by Pop~III and normal Pop~II stars at the same level, or even predominantly by Pop~II stars.

\section{Conclusions}

\label{conclu}
From comparing our model results with data, we propose simple diagnostics that exploit measured carbon and iron abundances to identify which metal-poor halo stars are the descendants of Pop~III stars ($>50\%$ of the metals in their ISM of formation) and to estimate the level of imprint from normal Pop~II SNe. These are our key findings:
\begin{itemize}
    \item CEMP-no stars with [C/Fe] $\ge +2.5$ are second-generation stars {\it solely} imprinted by massive first stars exploding with low to normal energy ($E_{SN}\leq 1.2\times 10^{51}$~erg). %This result is also probed by the abundances of the other chemical elements.
    \item CEMP-no stars with $\rm[C/Fe] <+2.5$ are likely Pop~III star descendants, which have been mainly polluted by Pop~III SNe with a non-negligible $\sim (10-30)\%$ contribution from Pop~II stars.
    \item The majority of C-normal stars have been enriched by normal Pop~II stars at $\gtrsim50\%$ level, and hence they do not show strong chemical features left by Pop~III SNe.
\end{itemize}

Conversely, our models show that Pop~III SNe can also enrich the ISM to low carbon values. 
%The majority of descendants with $\rm[C/Fe]\lesssim 0$ are predicted to be enriched by Pop~III SNe exploding with high-energy ($E_{SN}\geq 3\times 10^{51}$~erg). 
\ire{The [C/Fe] values of the pure descendants of Pop~III SNe exploding with high-energy ($E_{SN}\geq 3\times 10^{51}$~erg) are mostly located at [C/Fe] $\lesssim 0$.}
Therefore, we expect to find rare C-normal stars that are direct descendants of these very energetic Pop~III SNe, thus showing peculiar chemical abundance patterns. This is consistent with the recent discoveries of primordial hypernova descendants \citep[see][]{Skuladottir2021,Placco2021}.

Ultimately, our simple and general parametric model can be applied to interpret the chemical abundances measured in both ancient stars and in high-redshift absorption systems (see Salvadori et al. in this volume). According to our predictions, a C-excess measured in gaseous environments at high redshifts ($z > 4$) is a solid proof of chemical enrichment driven by Pop~III stars. 

The observed scatter of chemical abundance ratios is another key diagnostic. Our model shows a larger scatter with a higher fraction of Pop~III contribution. This is supported by observations of metal-poor stars in the Galactic halo (Fig.~\ref{fig:Yong_scarti}), and we predict that this is also true in the ISM observed at high redshifts. 

By using a powerful combination of models and data, we have exploited C and Fe abundances to identify first star descendants and quantify their Pop~III enrichment. However, using only these two elements is not sufficient to break degeneracies between different models of Pop~III SNe (mass and energy). In an upcoming study, we will thus expand the work presented here to include all the chemical elements available, exploiting to the fullest the very metal-poor stars to better understand the first stars in the Universe.
 
\begin{acknowledgements}
This project has received funding from the European Research Council (ERC) under the European Union’s Horizon 2020 research and innovation programme (grant agreement No 804240). I.V. and S.S. acknowledge support from the PRIN-MIUR17, prot. n. 2017T4ARJ5.
\end{acknowledgements}

\bibliographystyle{aa}
\bibliography{Main}

\end{document}